# Direct photoluminescence probing of ferromagnetism in monolayer two-dimensional CrBr$_3$

*Zhaowei Zhang,[†,∇] Jingzhi Shang,[†,‡,∇] Chongyun Jiang,[†,∇] Abdullah Rasmita,[†] Weibo Gao[\*,†,∥] and Ting Yu[\*,†]*

[†]Division of Physics and Applied Physics, School of Physical and Mathematical Sciences, Nanyang Technological University, Singapore 637371, Singapore

[‡]Institute of Flexible Electronics, Northwestern Polytechnical University, 127 West Youyi Road, Xi'an 710072, China

[∥]The Photonics Institute and Centre for Disruptive Photonic Technologies, Nanyang Technological University, 637371 Singapore, Singapore

**Abstract:** Atomically thin magnets are the key element to build up spintronics based on two-dimensional materials. The surface nature of two-dimensional ferromagnet opens up opportunities to improve the device performance efficiently. Here, we report the intrinsic ferromagnetism in atomically thin monolayer CrBr$_3$, directly probed by polarization resolved magneto-photoluminescence. The spontaneous magnetization persists in monolayer CrBr$_3$ with a Curie temperature of 34 K. The development of magnons by the thermal excitation is in line with the spin-wave theory. We attribute the layer-number dependent hysteresis loops in thick layers to the magnetic domain structures. As a stable monolayer material in air, CrBr$_3$ provides a convenient platform for fundamental physics and pushes the potential applications of the two-dimensional ferromagnetism.

Ferromagnetism in atomically thin magnet has been studied in a variety of van der Waals materials [1,2], including metallic Fe$_3$GeTe$_2$ [3,4], semiconducting Cr$_2$Ge$_2$Te$_6$ [5] and insulating CrI$_3$ [6]. Even though the long-range magnetic order is highly suppressed by the thermal excitation of magnons in a two-dimensional (2D) magnet at finite temperature [7], the magnetic anisotropy opens an energy gap in the magnon spectra and therefore, protects the ferromagnetism in two dimensions. The magnon-magnon interaction in such van der Waals ferromagnets also provides a platform to study the fundamental topological spin excitation, for example, Dirac magnon [8] and topological magnon surface state [9]. Moreover, in contrast to the three-dimensional ferromagnet, magnetic 2D materials show tunable magnetic properties due to their surface nature [1-3, 10-13]. Particularly the layer-number dependent [4, 6, 14] and gate-tunable magnetism [3, 10-13] opens a new way to build spintronic devices with high accuracy and efficiency [15-20].

Among various van der Waals ferromagnets, CrBr$_3$ is an interesting platform to study the magnetism in low dimensions and light matter interactions in magnetic materials. The neutron scattering has revealed the Dirac points in bulk CrBr$_3$ [21, 22], formed by acoustic and optical spin-wave modes, where both intralayer and interlayer exchange interactions play an important role. On the other hand, optical absorption spectra in CrBr$_3$ have shown the out-of-plane magnetic field dependence [23], suggesting potential applications in optoelectronics. However, magnetism in atomically thin CrBr$_3$, especially in monolayer limit, is still unknown.

In this work, we demonstrate the ferromagnetism in 2D van der Waals CrBr$_3$. We show the spontaneous magnetization in monolayer CrBr$_3$, probed by d-d transition induced photoluminescence (PL) with a polarization-resolved optical confocal setup. The magnon excitation is limited at the low temperature region, but shows an exponential development as further increasing the temperature, which is in line with the spin-wave theory. It is worthy to mention that CrBr$_3$ is much more stable in air as compared to CrI$_3$ as reported previously [6, 24], providing a convenient platform for magnetic material applications. Our study also shows the ferromagnetic interlayer coupling and magnetic domain induced hysteresis loops in multilayers, providing an opportunity to use magnetic domains as the information carrier in a van der Waals magnet.

The atomic structure of monolayer CrBr$_3$ is shown in Figure 1a. Cr$^{3+}$ ions are arranged in a honeycomb lattice, and the Cr-Br-Cr bond forms an angle of 95.1 degrees, suggesting that ferromagnetic superexchange interaction is energetically favorable. To study the magnetism in two dimensions, CrBr$_3$ flakes were exfoliated on a quartz substrate as shown in the insert of Figure 1b. Although in previous studies, reflection and absorption spectra have indicated the presence of $^4T_2$ parity forbidden d-d transition [23] as shown in Figure S1, we first experimentally uncovered the d-d transition induced PL at 1.35 eV (Figure 1b), excited by a continuous wave (CW) laser at 1.77 eV. All measurements were performed at 2.7 K, unless otherwise specified. We studied the laser power dependent PL under the same polarization configuration (Figure S2). The PL intensity scales linearly with the laser power. This linear dependence rules out the possibility that the PL arises from the defect-bond excitons whose PL intensity trends to saturate while increasing the laser power [25]. We further examined the PL spectra for various layer-thicknesses as shown in Figure 1b. The PL peak energy almost doesn't change for the thickness ranging from 6 to 73 nm, which suggests a localized transition, in agreement with the d-d transition as an inter-atom transition. According to Laporte rule, d-d transition is parity forbidden. To relax the Laporte rule, symmetry breaking must be introduced, such as spin-orbit coupling, Jahn-Teller distortion and the formation of odd-parity phonons [26]. The broad PL linewidth serves as the evidence for the strong vibrionic coupling, resulting in photon sidebands.

The d-d transition in the out-of-plane magnetic field shows circularly selective PL. Figure 1d shows the PL in -0.5 T and 0.5 T with $\sigma^+\sigma^+$ and $\sigma^-\sigma^-$ excitation-collection configurations, where $\sigma^+(\sigma^-)$ represents the left (right) circularly polarized light. The PL at $\pm 0.5$ T shows opposite helicity, indicating that the spin of electrons in CrBr$_3$ is coupled to the circularly polarized light.

The thickness of monolayer CrBr$_3$ is around 1 nm, determined by atomic force microscope (AFM) (Figure S3). To precisely detect the magneto-PL in a single magnetic domain, the PL emission was collected by a single-mode optical fiber and detected by an avalanche photodiode (APD). Figure 2a shows the polarization as a function of the magnetic field under $\sigma^+\sigma^+$, $\sigma^-\sigma^-$, $\sigma^+\sigma^-$, and $\sigma^-\sigma^+$ configurations. The polarization calculated by $\rho = \frac{I-(I_\uparrow+I_\downarrow)/2}{(I_\uparrow+I_\downarrow)/2}$ is proportional to the magnetic moments, where $I$ is the PL intensity recorded by APD while sweeping the magnetic field, and $I_\uparrow$ ($I_\downarrow$) is the PL intensity for fully spin up (down) states. The green symbols show the evolution of the PL intensity with the magnetic field sweeping from 0.1 to -0.1 T, and the orange symbols show the time reversal process.

We first discuss the polarization resolved PL as a function of the magnetic field. There are three features in the Figure 2a: (i) The polarization abruptly increases or decreases within a narrow field range; (ii) Except the points near the transition field, the polarization is almost independent to the magnetic field; (iii) The hysteresis loops have the same shape with the same excitation polarization, and it is independent to the polarization of the collection. We attribute these observations to the circularly selective absorption as shown in Figure 2b. Taking $\sigma^+\sigma^+$ configuration as an example, at 0.1 T the electrons with up-spin selectively absorb the $\sigma^+$ light and are excited to the upper states. As the magnetic field is swept to a negatively large point, thereby flipping the spin, less electrons can be excited by the $\sigma^+$ light. Therefore, the polarization suddenly drops. And in the time reversal process, the polarization shows a rapid increase at a certain field. The collection-polarization independent hysteresis loop is assigned to the depolarization of electrons at the excited state due to the electron-phonon scattering. This is in consistence with the strong vibronic coupling in the d-d transition. This electric dipole transition provides a way to optically probe the magnetic state of the 2D $CrBr_3$ [26].

The non-zero polarization at zero magnetic field (0 T) indicates the presence of spontaneous magnetization in monolayer $CrBr_3$. To further confirm the intrinsic ferromagnetism, we measured the hysteresis loops with various laser powers ranging from 10 to 100 $\mu W$ (Figure S4). Even though the APD count increases with the excitation power, the polarization as a function of the magnetic field doesn't change much. The laser power independent hysteresis loops rule out the effect of thermal excitation on the magnetization. We also studied the magneto-optical Kerr effect (MOKE) in the monolayer $CrBr_3$ on the Si/SiO substrate as shown in Figure S5(a), which agrees with magnetic hysteresis loops probed by PL.

The ferromagnetism in monolayer results from the Cr-Br-Cr superexchange interaction. In monolayer $CrBr_3$, six $Cr^{3+}$ ions form a honeycomb structure and each $Cr^{3+}$ ion is surrounded by six $Br^-$, forming an octahedral environment (Figure 1a). In this crystal field, the degeneracy of d orbit of Cr atom is lifted and the d level splits into $t_{2g}$ and $e_g$ bands. Three spin polarized electrons occupy the $t_{2g}$ band according to Hund's first rule. Therefore, the magnetic moment each $Cr^{3+}$ ion yields is ~$3\mu_B$, corresponding to the polarization of 0.175 in monolayer case. The magnetic moments of monolayer CrBr$_3$ align in the out-of-plane direction. The XXZ spin Hamiltonian [27] is adopted to describe the 2D ferromagnet: $H_{spin} = -A\sum_i (S_i^z)^2 - J\sum_{i,j} S_i S_j - \lambda \sum_{i,j} S_i^z S_j^z$, where $A$ is the single-ion anisotropy term, $J$ is the Heisenberg exchange term and $\lambda$ is the anisotropic exchange term. The quenched d-orbit results in a negligible single-ion anisotropy term and thereby $A$ is almost zero. The angle of the Cr-Br-Cr bond is around 90 degrees, which favors a ferromagnetic intralayer coupling and $J > 0$. The out-of-plane magnetic anisotropy corresponds to $\lambda > 0$.

The Curie temperature was experimentally determined by measuring the hysteresis loops at various temperatures (Figure 2c). When the temperature was increased above the Curie temperature at 34 K, a ferromagnetism-to-paramagnetism phase transition occurred. The $T_C$ of 34 K is only slightly lower than that of the bulk (*i.e.* 37 K). Figure 3 shows the polarization as a function of the temperature. The excitation of magnon by thermal fluctuation degrades the long-range magnetic order. We describe the reduced polarization as increasing the temperature within a spin-wave theory [27]. In an isotropic 2D spin system, the gapless magnon spectra leads to the absence of the spontaneous magnetization at finite temperature. Nevertheless, $\Delta_0$, the spin wave gap opens by the anisotropic energy, protects the long-range magnetic order, which plays an essential role in the 2D ferromagnet. The magnetization in units of $\hbar$ per Cr atom as a function of the temperature is described as: $M(T) = S - \frac{k_B T}{2\pi J S} e^{-\Delta_0/k_B T}$, where $S = 3/2$, and $k_B$ is Boltzmann constant [27]. The polarization $\rho$ is proportional to the $M$: $\rho(T) \sim S - \frac{k_B T}{2\pi J S} e^{-\Delta_0/k_B T}$. The solid line in Figure 3 shows the fitting results with this model, in line with our experimental data. In the low temperature region (< 15K), the polarization weakly depends on the temperature. Further increasing the temperature leads to an exponential development of magnons and the polarization rapidly decreases until vanishing at the $T_C$.

Next, we study the layer-layer interaction in 2D van der Waals $CrBr_3$. The interlayer coupling is revealed by a bilayer $CrBr_3$ with the thickness of about 2 nm, determined by AFM (Figure S3). Distinct from bilayer $CrI_3$, whose interlayer coupling at the ground states was antiferromagnetic [6], bilayer $CrBr_3$ preserves ferromagnetism as shown in Figure 4(a), suggesting a ferromagnetic interlayer coupling. Note that the transition field for bilayer $CrBr_3$ from fully spin up states to fully spin down states, is almost one order smaller than that for bilayer $CrI_3$ [6]. This is in line with the different anisotropic energies of these two ferromagnetic insulators.

Finally, we discuss the magnetism in multilayer $CrBr_3$. Figure 4b shows the polarization as a function of the magnetic field for 8, 54 and 73 nm samples. Different from the thin sample, whose rectangular hysteresis loops indicate single-domain and fully out-of-plane anisotropic magnetic ordering, the polarization of thicker $CrBr_3$ samples vanishes at 0 T. As the magnetic field was increased, the polarization became saturated after reaching a transition field, being similar to that of previously reported multilayer $Fe_3GeTe_2$ [4] and Co/Pt thin films [28]. We attribute this magnetic behavior to the formation of strip- and honeycomb-like magnetic domain structures as reported in Ref [29], which is beyond the resolution of the PL setup. The spot size of our excitation laser is about 1 $\mu$m, and thereby the polarization is contributed by several domains. As a result, the polarization at 0 T vanishes and shows a gradual evolution as increasing the magnetic field, and abruptly saturates at a certain magnetic field. The possible reason for forming this kind of domains might be the low out-of-plane anisotropic energy in thick $CrBr_3$ [30]. Even in a very low temperature (2.5K), the ratio of out-of-plane anisotropy to the exchange interaction is strongly dependent on the dimension. The enhanced out-of-plane anisotropy in thin layers might result from a reduced screening effect [24]. Even though we could expect a higher $T_C$ with a larger out-of-plane anisotropy, the competition of the increased thermal fluctuation in thin layers eventually makes the $T_C$ slightly lower than that of bulk.

To conclude, we demonstrate ferromagnetism in atomically thin $CrBr_3$ through polarization resolved magneto-PL. In monolayer $CrBr_3$, a rectangular hysteresis loop shows the spontaneous magnetization persists despite the thermal fluctuation. The polarization vanishes at 34 K due to the excitation of magnons. We also reveal the ferromagnetic interlayer coupling in a bilayer. Finally, the hysteresis loops of thick layers are assigned to the formation of strip- and honeycomb-like magnetic domains, whose magnetization is strongly dependent on the layer number of $CrBr_3$. Our study uncovers the magnetism in 2D $CrBr_3$ and might pave a way for novel opto-electronic and spintronic devices.

**Methods**

We fabricated atomically thin $CrBr_3$ layers by mechanical exfoliation on a quartz substrate in an argon-filled glovebox. The thickness of the sample was firstly estimated by optical contrast and then confirmed by AFM. After fabrication, the sample was load into a magneto-cryostat (Cryomagnetics close-cycle cryostat) with an out-of-plane magnetic field ranging from -7 T to 7 T. We use a home-made fiber based confocal microscope to perform the polarization resolved PL. Polarizers and waveplates are equipped on the excitation and collection arms to selectively excite and detect circularly polarized light. The PL spectra are obtained by a spectrometer (Andor Shamrock) with a CCD detector. To measure the hysteresis loops, the PL emission was collected by a single-mode fiber and detected by an APD.

**ASSOCIATED CONTENT**

**Supporting Information**

The Supporting Information is available free of charge on the ACS Publications website.

Additional details on configurational coordinate diagram for the $^4A_2$ to $^4T_2$ d-d transition, PL intensity as a function of the excitation laser power with a linear fitting, AFM images for monolayer and bilayer $CrBr_3$, hysteresis loops at various excitation laser powers, MOKE for two-dimensional $CrBr_3$ and temperature dependent hysteresis loops for bilayer and 73 nm $CrBr_3$.

**AUTHOR INFORMATION**

**Corresponding Authors**

*Email: wbgao@ntu.edu.sg

*Email: yuting@ntu.edu.sg*Email: wbgao@ntu.edu.sg

*Email: yuting@ntu.edu.sg

**Author Contributions**

$^\nabla$Z.Z., J.S. and C.J. contributed equally.

**ACKNOWLEDGMENTS**

We acknowledge the support from the Singapore National Research Foundation through a Singapore 2015 NRF fellowship grant (NRF-NRFF2015-03) and its Competitive Research Program (CRP Award No. NRF-CRP14-2014-02), Singapore Ministry of Education (MOE2016-T2-2-077 and MOE2016-T2-1-163), A*Star QTE programme and a start-up grant (M4081441) from Nanyang Technological University.

Figures:

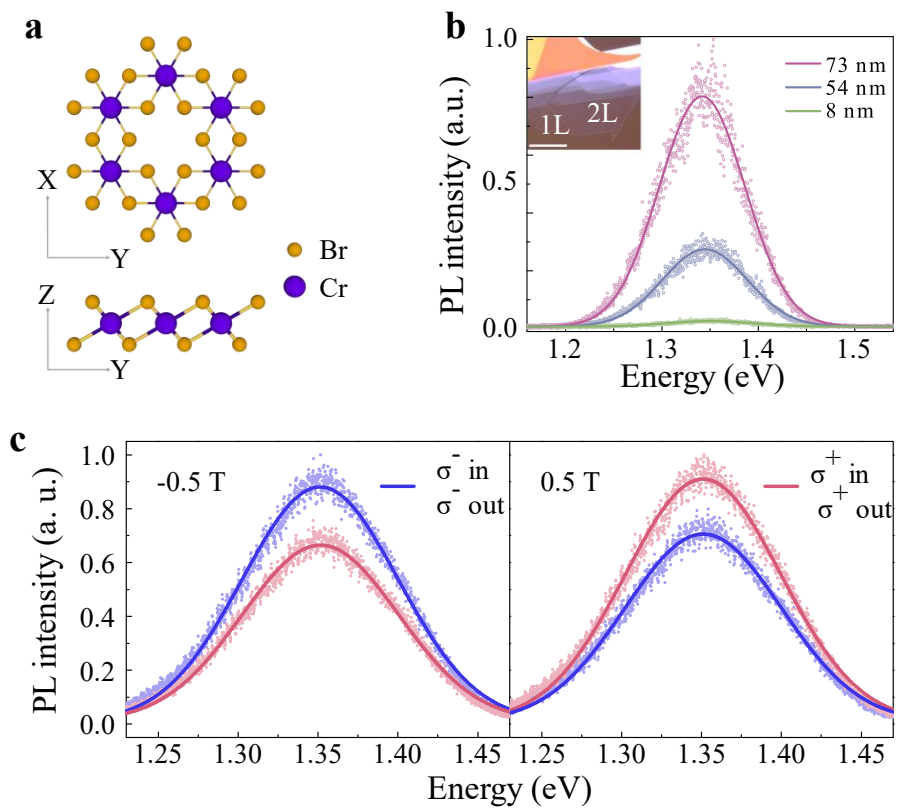

**Figure 1. Crystal structure and PL of two-dimensional van der Waals CrBr₃. a,** Top view and side view of the atomic structure of monolayer CrBr₃. The $Cr^{3+}$ ion was surrounded by six $Br^-$ ions, forming an octahedral environment. The Cr-Br-Cr bond forms an angle of 95.1 degrees. **b,** PL spectra for CrBr₃ with various thickness. Insert: Optical image of the exfoliated 2D CrBr₃ on a quartz substrate. The scale bar is 15 $\mu$m. c, Polarization resolved PL for monolayer CrBr₃ at ±0.5 T. All PL data were fitted by Gaussian functions.

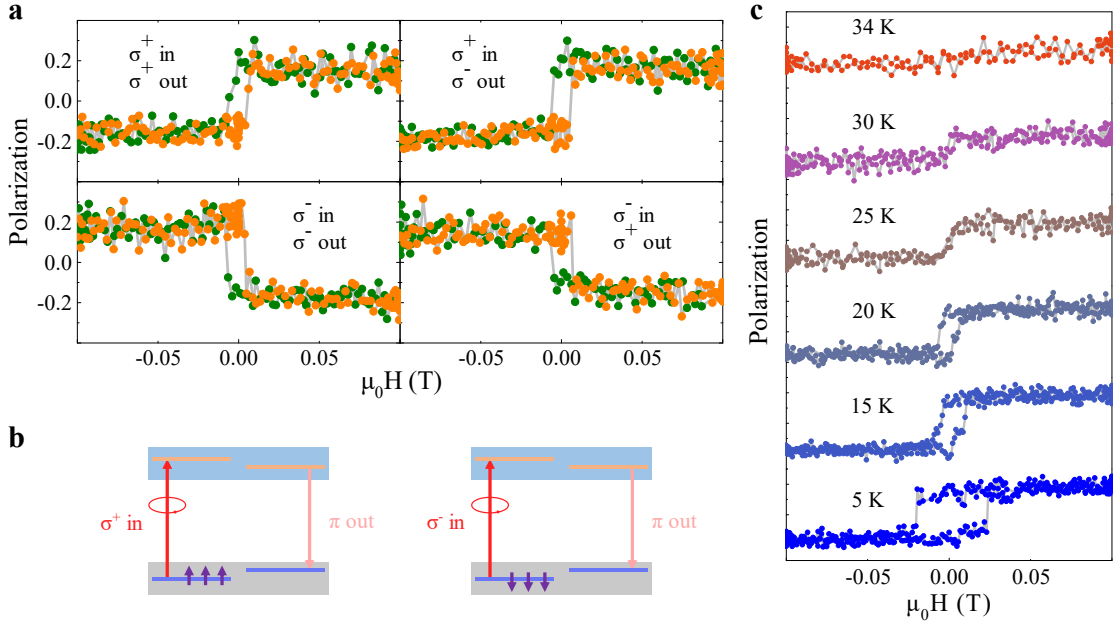

**Figure 2. Ferromagnetism in monolayer CrBr₃. a**, Polarization as a function of the magnetic field. The orange symbols show the polarization as the magnetic field is swept from -0.1 T to 0.1 T and the green symbols show the polarization as the magnetic field is swept from 0.1 T to -0.1 T. The none-zero polarization at zero magnetic field indicates the spontaneous magnetization. $\sigma^+(\sigma^-)$ is the left (right) circularly polarized light. **b**, Origin of the magnetic field dependent PL. The helicity of the absorption ties to the spin of the electrons at the ground states. The unbalance of the spin-up and spin-down states makes a higher $\sigma^+/\sigma^-$ absorption and eventually leads to the high/low PL emission. Due to the phonon scattering at the excited state, the output light is depolarized ($\pi$). c, Hysteresis loops at various temperature. The hysteresis loop disappears as the temperature was increased above the Curie temperature $T_C$ at 34 K, slight lower than that of the bulk crystal.

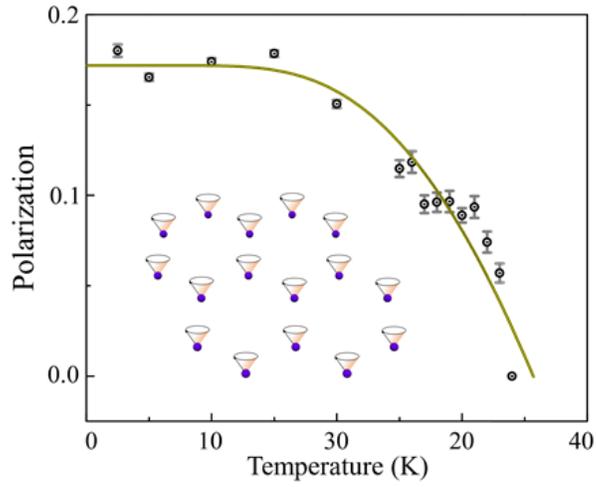

**Figure 3. The polarization as a function of temperature for monolayer CrBr₃.** The data is fitted by $\rho(T) \sim S - \frac{k_B T}{2\pi J S} e^{-\Delta_0/k_B T}$. Insert: Spin wave excited by the thermal fluctuation, which accounts for the decay of the polarization as increasing the temperature.

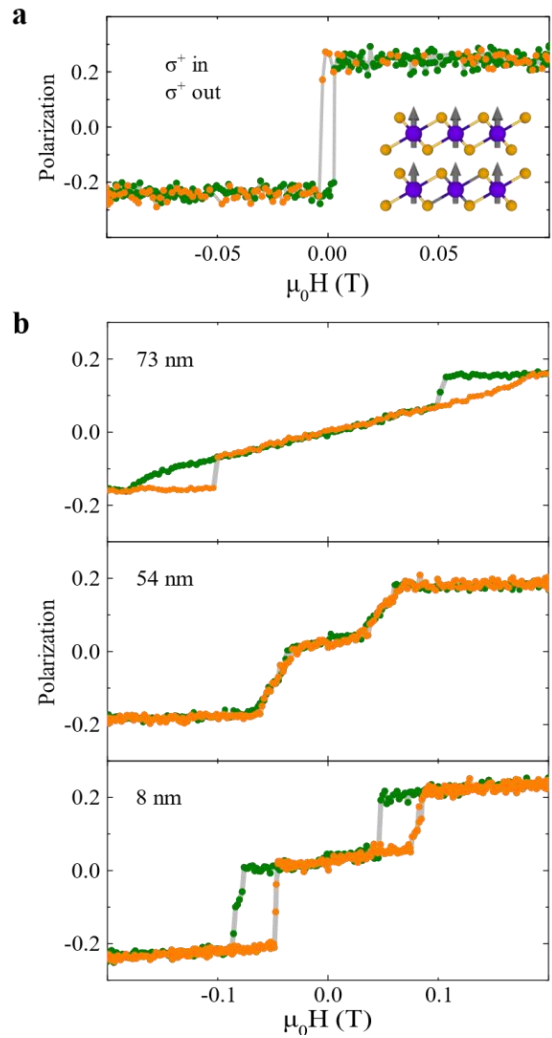

**Figure 4. Layer dependent magnetism in CrBr$_3$. a.** The hysteresis loop for bilayer CrBr$_3$. A non-zero polarization at zero magnetic field indicates a ferromagnetic interlayer coupling. **b.** The polarization as a function of the magnetic field for 8 nm, 54 nm and 73 nm CrBr$_3$. The formation of the strip- or honeycomb-like magnetic domains may account for the layer dependent magnetism in CrBr$_3$.

30. Jagla, E. A. Hysteresis loops of magnetic thin films with perpendicular anisotropy. *Phys. Rev. B* **2005**, *72*, 094406.

# Supporting Information for "Direct photoluminescence probing of ferromagnetism in monolayer two-dimensional material CrBr$_3$"

*Zhaowei Zhang,[†,∇] Jingzhi Shang,[†,‡,∇] Chongyun Jiang,[†,∇] Abdullah Rasmita,[†] Weibo Gao[\*,†,∥] and Ting Yu[\*,†]*

[†]Division of Physics and Applied Physics, School of Physical and Mathematical Sciences, Nanyang Technological University, Singapore 637371, Singapore

[‡]Institute of Flexible Electronics, Northwestern Polytechnical University, 127 West Youyi Road, Xi'an 710072, China

[∥]The Photonics Institute and Centre for Disruptive Photonic Technologies, Nanyang Technological University, 637371 Singapore, Singapore

[∇]These authors contributed equally.

\*Correspondence should be addressed to: wbgao@ntu.edu.sg or yuting@ntu.edu.sg

**S1**. The origin of the photoluminescence

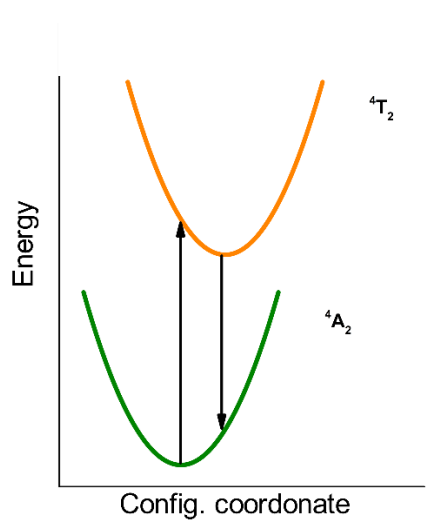

**Figure S1. Configurational coordinate diagram for the $^4A_2$ to $^4T_2$ d-d transition.** We assign the photoluminescence (PL) at 1.35 eV to the d-d transition in $CrBr_3$. We note the Ref[1] reported the absorption peak at 1.67 eV, which was assigned to the absorption to the $^4T_2$ state at 1.5 K. A Stokes shift of 320 meV between absorption and PL peaks is due to the strong electron-lattice coupling.

**S2**. Dependence of photoluminescence on the excitation power

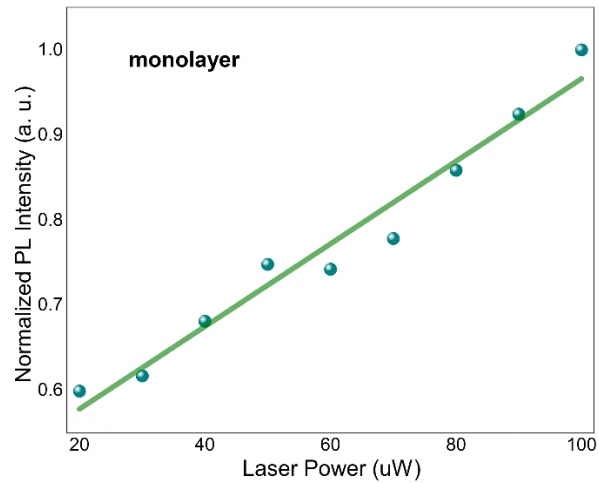

**Figure S2. PL intensity as a function of the excitation laser power with a linear fitting.** The PL intensity as a function of the laser power for monolayer is used to rule out the effect of defect on PL. The PL emission was collected by a single-mode fibre and detected by the APD. The linear dependence indicates that the PL doesn't arise from the defect-bound excitons [2], whose PL intensity trends to saturate at a high excitation power.

**S3.** AFM images for monolayer and bilayer CrBr$_3$

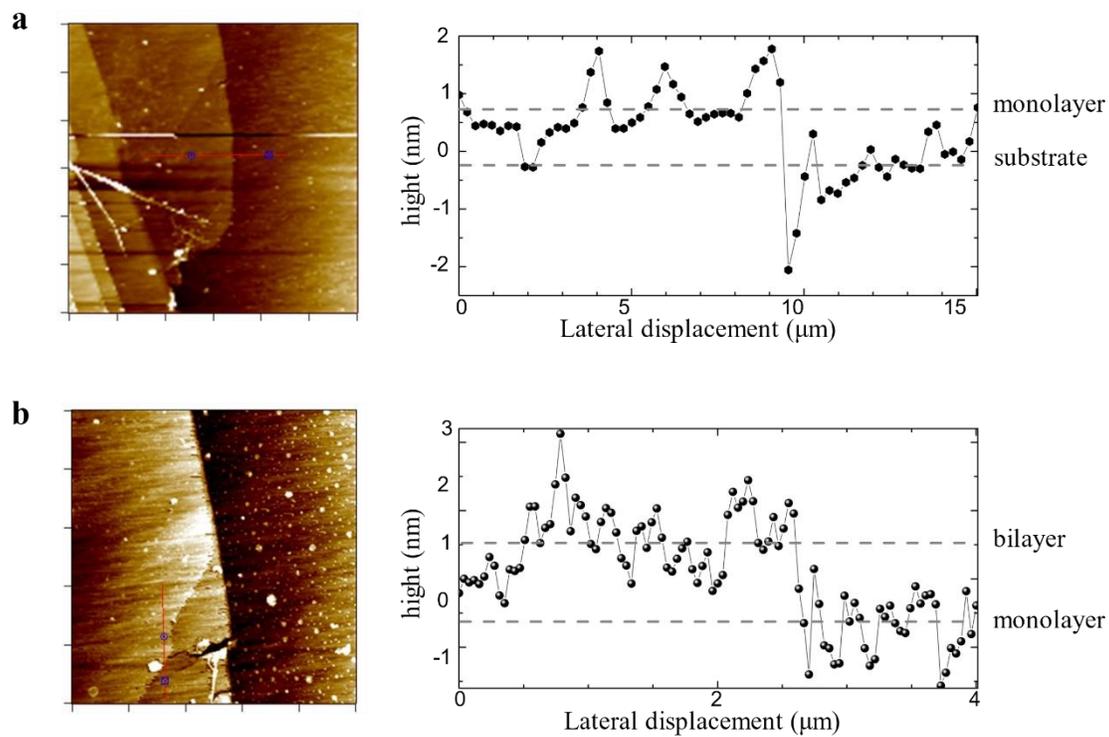

**Figure S3. AFM images for monolayer and bilayer CrBr$_3$.** The thickness of the monolayer and bilayer was first examined by the optical contrast and confirmed by AFM. AFM images show the thicknesses of monolayer and bilayer are around 1 and 2 nm, respectively.

**S4.** Hysteresis loops at various excitation laser powers

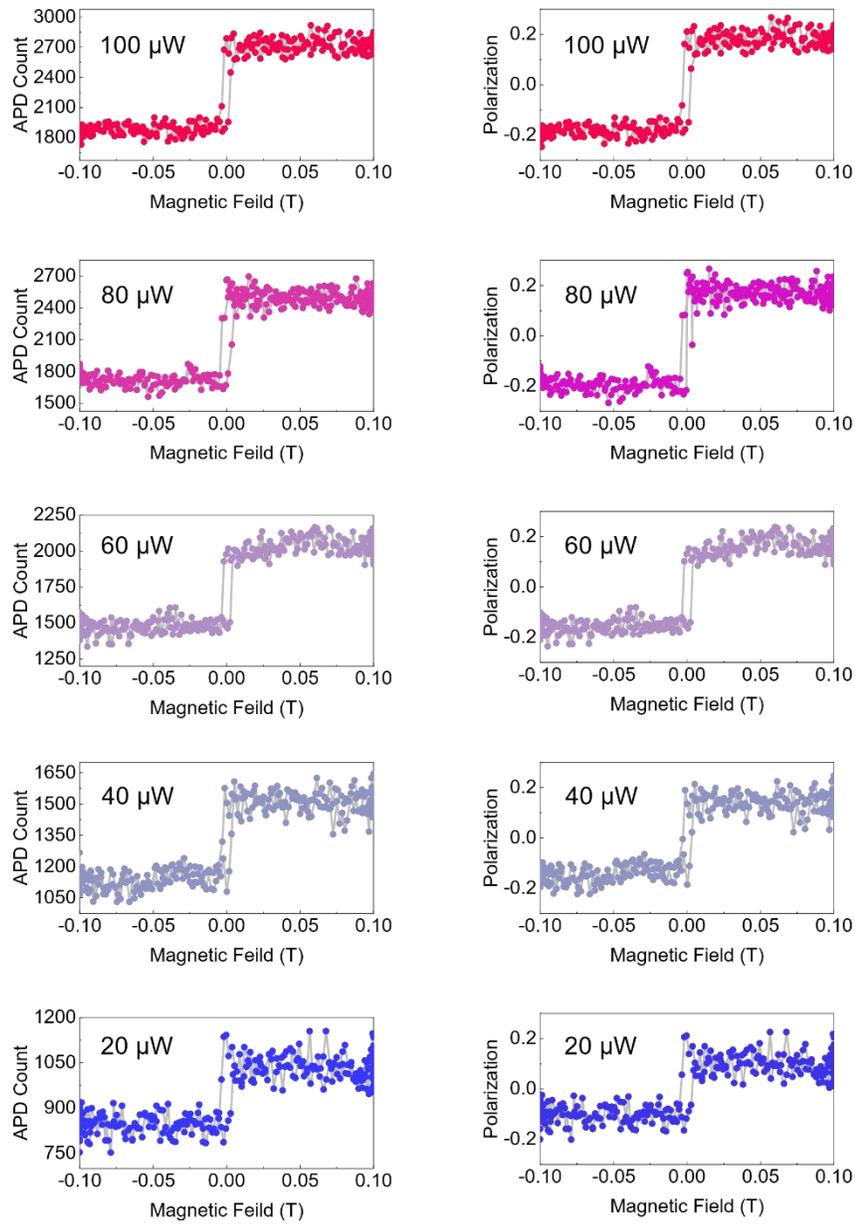

**Figure S4. Hysteresis loops at various excitation laser powers.** The left column shows the APD count as a function of the magnetic field under $\sigma^+\sigma^+$ configuration, and the left column shows the calculated polarization as a function of the magnetic field. By studying the laser power dependent hysteresis loops, we rule out the effect of thermal excitation on magnetism.

Furthermore, the polarization is almost independent to the laser power, indicating that the defined polarization is proportional to the magnetization.

**S5**. Magneto-optical Kerr effect (MOKE) for CrBr$_3$

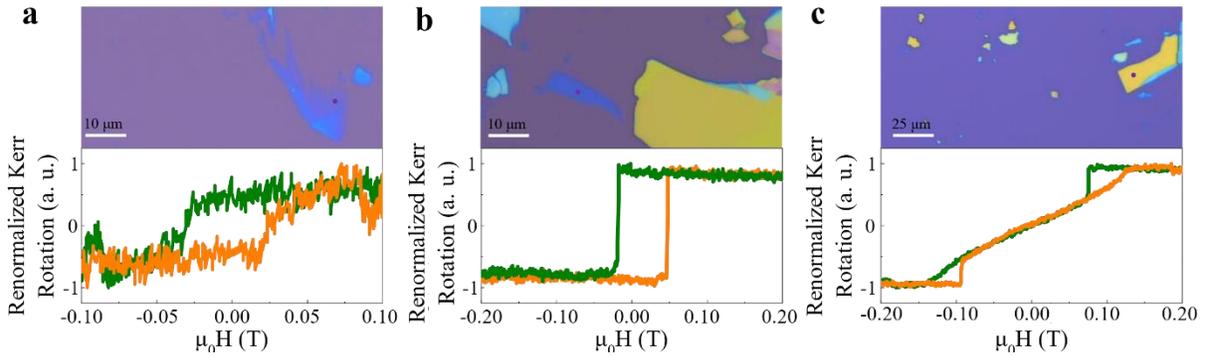

**Figure S5. MOKE for two-dimensional CrBr$_3$.** To further confirm the ferromagnetism in two-dimensional CrBr$_3$, we also performed MOKE measurements. (a), (b) and (c) show MOKE for CrBr$_3$ with various thickness. CrBr$_3$ flakes were exfoliated on Si/SiO$_2$ substrates and then loaded into a magneto-cryostat (Cryomagnetics close-cycle cryostat). The MOKE measuring protocols were described in Ref [3] . The magnetic field was applied in the out-of-plane direction. The incident laser with the wavelength of 405 nm was adopted to perform the MOKE measurement and the laser power was 12 $\mu W$. The MOKE results agree well with the magnetic hysteresis loops probed by PL, indicating that the direct PL probing ferromagnetism is reliable.

## S6. $T_C$ for bilayer and bulk $CrBr_3$

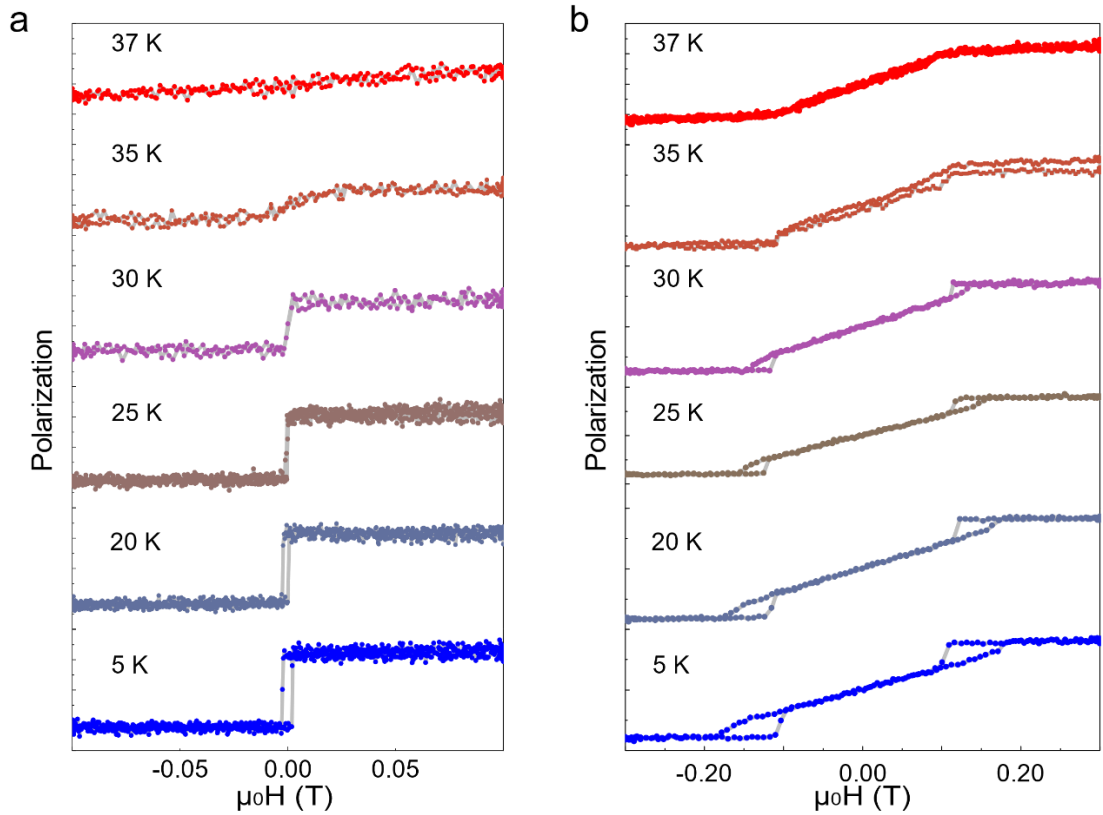

**Figure S6. Temperature dependent hysteresis loops for bilayer and 73 nm $CrBr_3$.** The Curie temperatures of the bilayer and 73 nm $CrBr_3$ were determined by increasing the temperature until the hysteresis loops disappeared. The $T_C$ for bilayer is around 35 K. The $T_C$ for 73 nm sample is around 37 K, the same as bulk $CrBr_3$ as reported previously[1].